\documentclass{wscpaperproc}
\usepackage{latexsym}
\usepackage{graphicx}
\usepackage{mathptmx}
\usepackage[usenames,dvipsnames]{color}

\usepackage{amsmath}
\usepackage{amsfonts}
\usepackage{amssymb}
\usepackage{amsbsy}
\usepackage{amsthm}

\usepackage[pdftex,colorlinks=true,urlcolor=blue,citecolor=black,anchorcolor=black,linkcolor=black]{hyperref}

\newtheoremstyle{wsc}
{3pt}
{3pt}
{}
{}
{\bf}
{}
{.5em}
{}

\theoremstyle{wsc}

\begin{document}

%
%
\WSCpagesetup{Rhee and Glynn}

\title{A NEW APPROACH TO UNBIASED ESTIMATION FOR SDE'S}

\author{Chang-han Rhee\\
Peter W. Glynn\\[12pt]
Stanford University\\
Stanford, CA 94305, U.S.A.
}

\maketitle

\section*{ABSTRACT}
In this paper, we introduce a new approach to constructing unbiased estimators when computing expectations of path functionals associated with stochastic differential equations (SDEs). Our randomization idea is closely related to multi-level Monte Carlo and provides a simple mechanism for constructing a finite variance unbiased estimator with ``square root convergence rate" whenever one has available a scheme that produces strong error of order greater than 1/2 for the path functional under consideration. 

\section{INTRODUCTION}
\label{sec:intro}

We have recently developed a general approach to constructing unbiased estimators, given a family of biased estimators. It turns out that the conditions guaranteeing its validity are closely related to those associated with multi-level Monte Carlo methods; see Rhee and Glynn (2012) for details and a more complete discussion of the theory. In this paper, we briefly describe the idea in the setting of computing solutions of stochastic differential equations and provide an initial numerical exploration intended to shed light on the method's potential effectiveness. As we will see below, the conditions under which our estimator produces an algorithm with ``square root convergence rate" essentially coincide with the conditions required by multi-level Monte Carlo to converge at the same rate.

In particular, suppose that we wish to compute an expectation of the form $\alpha = E k(X)$, where $X =(X(t): t\geq0)$ is the solution to the SDE
\begin{align}
                       dX(t) = \mu(X(t))dt + \sigma(X(t))dB(t),                           \label{eq:1}
\end{align}
$B =(B(t):t\geq0)$ is m-dimensional standard Brownian motion, $k: C[0, \infty) \to R$, and $C[0, \infty)$ is the space of continuous functions mapping $[0,\infty)$ into $R^d$. In general, the random variable (rv) $k(X)$ can not be simulated exactly, because the underlying infinite-dimensional object $X$ can not be generated exactly. Instead, one typically approximates $X$ via a discrete-time approximation $X_h(\cdot)$. For example, the simplest such approximation is the Euler time-stepping algorithm given by
\begin{align}
                      X_h((k+1)h) = X_h(kh) + \mu(X_h(kh))h + \sigma(X_h(kh)) (B(k+1)h) - B(kh))               \label{eq:2}
\end{align}
that defines $X_h$ at the time points $0, h, 2h, ...,$ with $X_h$ defined at intermediate values via (for example) linear interpolation. Because (\ref{eq:2}) is only an approximation to the dynamics represented by (\ref{eq:1}), the rv $k(X_h)$ is only an approximation to $k(X)$, and consequently $k(X_h)$ is a biased estimator for the purpose of computing $\alpha$. The traditional means of dealing with this is to intelligently select the step size $h$ and number of independent replications $R$ as a function of the computational budget $c$, so as to maximize the rate of convergence. However, as pointed out by Duffie and Glynn (1995), such biased numerical schemes inevitably lead to Monte Carlo estimators for $\alpha$ that exhibit slower convergence rates than the ``canonical" order $c^{-1/2}$ rate associated with Monte Carlo in the presence of unbiased finite variance estimators.

However, several years ago, Giles (2008) introduced an intriguing multi-level idea to deal with such biased settings that can dramatically improve the rate of convergence and can even, in some settings, achieve the canonical ``square root" convergence rate associated with unbiased Monte Carlo. His approach does not construct an unbiased estimator, however. Rather, the idea is to construct a family of estimators (indexed by the desired error tolerance $\epsilon$) that has controlled bias. In this paper, we show how it is possible, in a similar computational setting, to go one step further and to produce (exactly) unbiased estimators. The remainder of this paper is organized as follows: We discuss the idea in Section 2 of this paper, while Section 3 is devoted to an initial computational exploration of this approach.

\section{THE BASIC IDEA}
We consider here a sequence $(X_{h_n}: n>=0)$ of discrete-time time-stepping approximations to $X$ that are all constructed on a common probability space in such a way that:
\begin{enumerate}
          \item[i)] $Ek(X_{h_n}) = Ek(X) + O(h_n)$ as $h_n \to 0$;

         \item[ii)] $E| k(X_{h_n}) - k(X) |^2 = O(h_n^{2r})$ as $h_n \to 0$
\end{enumerate}
for some $r >0$, where $O(f(n))$ represents a function which is bounded by some constant multiple of $f(\cdot)$ as $h_n \to 0$. Assuming, as is often the case for such discretization schemes, that the scheme generates normal rv's that are intended to mirror the Brownian increments of the process $B$ driving the SDE (as in the Euler scheme (1.2) above), the easiest way to algorithmically obtain an approximating sequence $X_{h_n}$ to $X$ in which the $X_{h_n}$'s are jointly defined on the same probability space is by successive binary refinement, so that $h_n = 2^{-n}$. In this setting, the new Brownian motion values ($B(j2^{-(n+1)})$: $j$ odd) required at discretization $2^{-(n+1)}$ can be obtained from the existing values ($B(j2^{-n}): j \geq0$) by generating $B((2k+1) 2^{-(n+1)})$ from its conditional distribution given $B(k2^{-n})$ and $B((k+1)2^{-n})$. On the other hand, one's ability to obtain i and ii depends both on the path functional $k$ and on one's choice of discretization scheme.

In particular, suppose that one has established that the discretization $X_h$ exhibits strong order $r$. This implies that
\begin{align*}
             E \sup\{ | X_h(kh) - X(kh) |^{2r} : 0\leq k \leq \lfloor t/h\rfloor \} = O(h^{2r}).
\end{align*}
Thus, if $k$ is (for example) a ``Lipschitz final value" expectation so that $k(x) = g(x(1))$ for some Lipschitz function $g: R^d \to R$, ii is satisfied. In addition, if $k$ is further assumed to be smooth with $| k(X) |$ integrable, then i is satisfied whenever the discretization $X_h$ is known to be of weak order 1 or higher. It should be noted that these conditions are (very) closely related to those that appear in the literature on multi-level Monte Carlo for SDEs.

Note that each of the $k(X_{2^{-n}})$'s is a biased estimator for $\alpha = E k(X)$. To obtain an unbiased estimator, observe that ii) implies the existence of $p > 0$ such that
\begin{align*}
            \sum_{n = 1}^\infty E  2^{np} | k(X_{2^{-n}}) - k(X_{2^{-(n-1)}}) | ^{2r} < \infty.
\end{align*}
Consequently,
\begin{align*}
            \sum_{ n=1}^\infty 2^{np} | k(X_{2^{-n}}) - k(X_{2^{-(n-1)}}) |^{2r} < \infty\quad a.s.
\end{align*}
from which it follows that
\begin{align*}
                               | k(X_{2^{-n}}) - k(X_{2^{-(n-1)}}) | = O(2^{-np})\quad a.s.
\end{align*}
as  $n \to \infty$, and hence (in view of ii),
\begin{align*}
                              k(X) = k(X_1) + \sum_{ n=1}^\infty  k(X_{2^{-n}}) - k(X_{2^{-(n-1)}})
\end{align*}

We now introduce a rv $N$, independent of $B$, that takes values in the positive integers and has a distribution with unbounded support (so that $P(N>n)>0$ for $n\geq1$). For such a rv $N$,
\begin{align*}
Ek(X) 
&= E k(X_1) + \sum_{n=1}^\infty E (k(X_{2^{-n}}) - k(X_{2^{-(n-1)}})) I(N \geq n)/ P(N \geq n)\\
&= E \left[k(X_1) + \sum _{n=1} ^N (k(X_{2^{-n}}) - k(X_{2^{-(n-1)}})) / P(N \geq n)\right]\\
&\triangleq E Z.
\end{align*}

Note that $Z$ is an unbiased estimator for $\alpha$. This suggests computing $\alpha$ by generating iid replicates of the rv $Z$. Of course, the ``square root" convergence rate of such an estimator is not guaranteed. Given the role that finiteness of the variance plays in obtaining such convergence rates, we next study this issue.

Set $\Delta_i = k(X_{2^{-i}}) - k(X_{2^{-(i-1)}})$ for $i \geq 1$ and observe that
\begin{align*}
E Z^2 
&=  E k^2 (X_1) + E \sum_{i=1}^N \Delta_i^2 / P(N\geq i)^2 + 2 E k(X_1) \sum_{i=1}^N \Delta_i / P(N \geq i) + 2 E \sum_{i=1}^N \sum_{j = i+1}^N  \Delta_i \Delta_j /(P(N\geq i) P(N\geq j))\\
&= E k^2 (X_1) + \sum_{i=1}^\infty E \Delta_i^2 / P(N\geq i) + 2E k(X_1) \sum_{i=1}^\infty \Delta_i + 2 \sum_{i=1}^\infty \sum_{j= i+1}^\infty E \Delta_i \Delta_j / P(N\geq i) \\
&= E k^2(X_1) + \sum_{i=1}^\infty E \Delta_i^2/P(N\geq i)+ 2 E k(X_1)(k(X)-k(X_1)) + 2 \sum_{i=1}^\infty E\Delta_i (k(X) - k(X_{2^{-i}}))/P(N\geq i)\\
&\leq E k^2(X_1) + 2 E k(X_1) (k(X) - k(X_1)) + \sum_{i=1}^\infty O(2^{-2ri})/P(N\geq i),
\end{align*}
so that $\text{var} Z < \infty$ if $P(N\geq i) \sim c 2^{- \gamma i}$ as $i \to \infty$, for $0 < \gamma < 2r$ (where $a_i \sim b_i$ means that $a_i/b_i \to 1$ as $i \to \infty$).

Finally, Glynn and Whitt (1992) prove that ``square root convergence rate"  ensues  if $\text{var} Z < \infty $ and if the expected computational effort required per replication of $Z$ is finite. The expected computational ``work" required for each $Z$ is (roughly) given by 
$$E \sum_{i=0}^N t_i,$$
where $t_i$ is the incremental effort required to compute $k(X_{2^{-i}})$ (given $k(X_1), \ldots , k(X_{2^{-(i-1)}})$), and hence can be expressed as 
\begin{align}
\sum_{i=0}^\infty t_i P(N\geq i). \label{eq:3}
\end{align}

An approximation to $t_i$ is $t_i = 2^{i-1} $ (the number of additional Gaussian rv's needed to generate $X_{2^{-i}}$). In order that (\ref{eq:3}) be finite, we require that $\gamma > 1$. Consequently, a square root convergence rate is ensured when $2r> 1$ ( in which case we can, for example, choose $\gamma = (1+2r)/2$).

\section{A PRELIMINARY COMPUTATIONAL INVESTIGATION}
In this section, we implement our method and compare it to the multilevel Monte Carlo algorithm suggested in Giles (2008). We consider two examples:\\

{\it Geometric Brownian Motion} (GBM): The process under consideration here is the solution to 
$$ dX(t) = rX(t) dt + \sigma X(t) dB(t)$$
subject to $X(0)=1$, $r=0.05$, $\sigma = 0.2$, the functional $k$ is $k(x) = x(1)$. For this set of parameters, 
$$E k(X) = 0.104506.$$

{\it Cox-Ingersoll-Ross process} (CIR): Here, $X$ solves 
$$ dX(t) = \kappa (\theta - X(t))dt + \sigma \sqrt{X(t)} dB(t),$$
subject to $X(0)= 0.04$, $\kappa = 5$, $\theta = 0.04$, and $\sigma = 0.25$. The functional $k$ is taken here to be $k(x) = x(1)$. For this example, 
$$E k(X) = 0.04.$$

The numerical scheme used to solve each of the above SDE's was the Milstein scheme; see Kloeden and Platen (1992). For the above problems, we expect $r=1$. For the purpose of this paper, we do not try to optimize the distribution of $N$, and instead choose $N$ so that 
$$ P(N \geq i) = 2^{3i/2}$$
for $i\geq 1$. (In other words, we choose $\gamma$ as the midpoint between $1$ and $2r$, although any choice in $(1,2)$ would provide "square root convergence rate".)

To compare our method to the multi-level Monte Carlo (MLMC) mehtod, we take the view (as in Giles, 2008) that the root mean square error (RMSE) $\epsilon$ to be achieved by the algorithm is given. Giles (2008) provides a complete description of how to construct a MLMC estimator achieving approximate RMSE $\epsilon$; we have implemented that version of MLMC here. For our unbiased estimator, we generate independent and identically distributed (iid) replications of the rv $Z$ until such time as the approximate RMSE is less than or equal to $\epsilon$. In other words, our estimator for $\alpha$ is 
\begin{align} 
\frac{1}{N(\epsilon)} \sum_{i=1}^{N(\epsilon)} Z_i, \label{eq:4}
\end{align}
where the $Z_i$'s are iid replicates of $Z$, and 
$$ N(\epsilon) = \inf \left\{ n \geq 1: \frac{1}{n(n-1)} \sum_{i=1}^n \left(Z_i - \frac{1}{n}\sum_{j=1}^N Z_j\right)^2 \leq \epsilon^2\right\}$$
is the first time that the sample RMSE of the sample mean drops below $\epsilon$. We use the stopping rule $N(\epsilon)$ in order to permit easy computation for our estimator, although its use is somewhat unnatural for our estimator (since its use induces bias in our estimator).

For each of our two examples, we provide two tables. The first table for each example concerns our new estimator (\ref{eq:4}); IRE stands for ``intended relative error", and $k\%$ then corresponds to  setting $\epsilon = (k/100) |\alpha|$. The 90\% confidence interval is then obtained by taking 100 replications of (4) for a given $\epsilon$, computing the sample mean and sample standard deviation of the 100 observations, and constructing a confidence interval based on the normal approximation. The column corresponding to RMSE is the square root of the average, over the 100 observations, of the square of (4) minus $E Z$. (Thus, RMSE is reporting the actual root mean square error of the estimator, rather than the intended RMSE that the estimator has been designed to attain asymptotically.) The final column, denoted ``work", reports a 90\% confidence interval for the expected number of normal rv's generated to construct (\ref{eq:4}), based on our 100 samples. The second table for each example provides a corresponding set of values for the MLMC estimator.

\begin{table}
\caption{Unbiased estimation for GBM\label{tab: UBGBM}}
\vspace{1mm}
\centering
\begin{tabular}{c|c|c|c}\hline
IRE & 90\% Confidence Interval & RMSE & Work \\\hline

25\%  & 0.108318  $\pm$  0.004503  & 0.02759  &      388.3  $\pm$        23.3  \\

10\%  & 0.105360  $\pm$  0.001874  & 0.01140  &     2850.8  $\pm$       145.7  \\

5\%  & 0.105244  $\pm$  0.000916  & 0.00560  &    11278.3  $\pm$       221.0  \\

2\%  & 0.104799  $\pm$  0.000388  & 0.00237  &    73451.9  $\pm$      1198.8  \\

1\%  & 0.104500  $\pm$  0.000174  & 0.00105  &   291488.1  $\pm$      2157.1  \\

0.5\%  & 0.104492  $\pm$  0.000088  & 0.00054  &  1170663.9  $\pm$      6067.8  \\
\hline
\end{tabular}
\end{table}

\begin{table}
\caption{MLMC estimation for GBM\label{tab: MLMCGBM}}
\vspace{1mm}
\centering
\begin{tabular}{c|c|c|c}\hline
IRE & 90\% Confidence Interval & RMSE & Work \\\hline

25\%  & 0.103753  $\pm$  0.002693  & 0.01635  &     3130.3  $\pm$         2.7  \\

10\%  & 0.104648  $\pm$  0.001179  & 0.00716  &     3842.5  $\pm$        10.5  \\

5\%  & 0.104908  $\pm$  0.000551  & 0.00337  &     6378.9  $\pm$        31.9  \\

2\%  & 0.104465  $\pm$  0.000218  & 0.00133  &    24901.3  $\pm$       196.0  \\

1\%  & 0.104293  $\pm$  0.000129  & 0.00081  &    93767.1  $\pm$       735.3  \\

0.5\%  & 0.104468  $\pm$  0.000060  & 0.00037  &   377783.8  $\pm$      4248.2  \\
\hline
\end{tabular}
\end{table}

\begin{table}
\caption{Unbiased estimation for CIR\label{tab: UBCIR}}
\vspace{1mm}
\centering
\begin{tabular}{c|c|c|c}\hline
IRE & 90\% Confidence Interval & RMSE & Work \\\hline

25\%  & 0.039914  $\pm$  0.001779  & 0.01080  &      883.1  $\pm$        61.4  \\

10\%  & 0.040272  $\pm$  0.000741  & 0.00450  &     5548.5  $\pm$       180.7  \\

5\%  & 0.040206  $\pm$  0.000342  & 0.00208  &    22693.3  $\pm$       690.8  \\

2\%  & 0.039880  $\pm$  0.000143  & 0.00087  &   142480.8  $\pm$      1790.4  \\

1\%  & 0.039953  $\pm$  0.000069  & 0.00042  &   563996.7  $\pm$      3609.4  \\

0.5\%  & 0.040022  $\pm$  0.000030  & 0.00018  &  2259485.0  $\pm$      8235.4  \\
\hline
\end{tabular}
\end{table}

\begin{table}
\caption{MLMC estimation for CIR\label{tab: MLMCCIR}}
\vspace{1mm}
\centering
\begin{tabular}{c|c|c|c}\hline
IRE & 90\% Confidence Interval & RMSE & Work \\\hline

25\%  & 0.040543  $\pm$  0.000973  & 0.00593  &     3731.2  $\pm$        13.1  \\

10\%  & 0.040186  $\pm$  0.000415  & 0.00252  &    11518.9  $\pm$        44.8  \\

5\%  & 0.039880  $\pm$  0.000208  & 0.00127  &    42503.8  $\pm$       139.4  \\

2\%  & 0.039862  $\pm$  0.000100  & 0.00062  &   265948.6  $\pm$       535.7  \\

1\%  & 0.040005  $\pm$  0.000044  & 0.00027  &  1098672.3  $\pm$      6866.5  \\

0.5\%  & 0.040008  $\pm$  0.000023  & 0.00014  &  4572949.2  $\pm$      5221.9  \\
\hline
\end{tabular}

\end{table}

Our results are reasonably comparable to those associated with MLMC, despite the fact that we have done essentially no tuning to optimize the distribution of $N$. In addition, our estimator is (arguably) easier to implement than MLMC, since (in its current form) there are no algorithmic parameters that are estimated ``on the fly" within the algorithm (in contrast to MLMC). Thus, the unbiased estimators introduced here offer a promising computational alternative to MLMC in the presence of SDE numerical schemes having a strong order greater than 1/2.

\section*{REFERENCES}
\begin{hangref}
\item Duffie, D. and P. W. Glynn. 1995. Efficient Monte Carlo simulation of security prices. {\it Annals of Applied Probability} 5(4):897-905.
\item Giles, M. B. 2008. Multilevel Monte Carlo path simulation. {\it Operations Research} 56(3):607–617.
\item Glynn, P. W. and W. Whitt. 1992. The asymptotic efficiency of simulation estimators. {\it Operations Research} 40:505-520
\item Kloeden, P.E. and E. Platen. 1992. {\it Numerical Solution of Stochastic Differential Equations}. Berlin: Springer-Verlag.
\item Rhee, C. and P. W. Glynn. 2012. From low bias to no bias: Application to SDE's. Submitted for publication.

\end{hangref}

\section*{AUTHOR BIOGRAPHIES}

\noindent {\bf CHANG-HAN RHEE} is currently a Ph.D. student in the Institute for Computational and Mathematical Engineering at
Stanford University. He graduated with B.Sc. in the Department of Mathematics and Department of Computer Science at Seoul National University, South Korea. His research interests include simulation, computational probability, sensitivity analysis and stochastic control. His email address is \email{chrhee@stanford.edu} and his web page is \url{http://www.stanford.edu/~chrhee/}.\\

\noindent {\bf PETER W. GLYNN} is the currently Chair of the Department of Management Science and Engineering at Stanford University. He received his Ph.D in Operations Research from Stanford University in 1982. He then joined the faculty of the University of Wisconsin at Madison, where he held a joint appointment between the Industrial Engineering Department and Mathematics Research Center, and courtesy appointments in Computer Science and Mathematics. In 1987, he returned to Stanford, where he joined the Department of Operations Research. He is now the Thomas Ford Professor of Engineering in the Department of Management Science and Engineering, and also holds a courtesy appointment in the Department of Electrical Engineering. From 1999 to 2005, he served as Deputy Chair of the Department of Management Science and Engineering, and was Director of Stanford's Institute for Computational and Mathematical Engineering from 2006 until 2010. He is a Fellow of INFORMS and a Fellow of the Institute of Mathematical Statistics, has been co-winner of Best Publication Awards from the INFORMS Simulation Society in 1993 and 2008, was a co-winner of the Best (Biannual) Publication Award from the INFORMS Applied Probability Society in 2009, and was the co-winner of the John von Neumann Theory Prize from INFORMS in 2010. In 2012, he was elected to the National Academy of Engineering. His research interests lie in simulation, computational probability, queueing theory, statistical inference for stochastic processes, and stochastic modeling. His email address is \email{glynn@stanford.edu} and his web page is \url{http://www.stanford.edu/~glynn/}.

\end{document}